\pdfoutput=1
\documentclass[cernpreprint, atlasdraft=false, coverpage=false, texlive=2023, USenglish, texmf, orcidlogo]{atlasdoc}
 
\usepackage{atlaspackage}
\usepackage[block=none,backref=false]{atlasbiblatex}
 
\usepackage[BSM=true]{atlasphysics}

\graphicspath{{logos/}{figures/}}

\newcommand{\ninogen}{\ensuremath{\mathchoice{\displaystyle\raise.4ex\hbox{\(\displaystyle\tilde\chi^0_{1,2,3,4}\)}}{\textstyle\raise.4ex\hbox{\(\textstyle\tilde\chi^0_{1,2,3,4}\)}}{\scriptstyle\raise.3ex\hbox{\(\scriptstyle\tilde\chi^0_{1,2,3,4}\)}}{\scriptscriptstyle\raise.3ex\hbox{\(\scriptscriptstyle\tilde\chi^0_{1,2,3,4}\)}}}\xspace}
\newcommand{\ninoggm}{\ensuremath{\mathchoice{\displaystyle\raise.4ex\hbox{\(\displaystyle\tilde\chi^0_{1,2}\)}}{\textstyle\raise.4ex\hbox{\(\textstyle\tilde\chi^0_{1,2}\)}}{\scriptstyle\raise.3ex\hbox{\(\scriptstyle\tilde\chi^0_{1,2}\)}}{\scriptscriptstyle\raise.3ex\hbox{\(\scriptscriptstyle\tilde\chi^0_{1,2}\)}}}\xspace}
\newcommand{\chinogen}{\ensuremath{\mathchoice{\displaystyle\raise.4ex\hbox{\(\displaystyle\tilde\chi^{\pm}_{1,2}\)}}{\textstyle\raise.4ex\hbox{\(\textstyle\tilde\chi^{\pm}_{1,2}\)}}{\scriptstyle\raise.3ex\hbox{\(\scriptstyle\tilde\chi^{\pm}_{1,2}\)}}{\scriptscriptstyle\raise.3ex\hbox{\(\scriptscriptstyle\tilde\chi^{\pm}_{1,2}\)}}}\xspace}

\AtlasTitle{A statistical combination of ATLAS Run 2 searches for charginos and neutralinos at the LHC }

\AtlasAbstract{
 
Statistical combinations of searches for charginos and neutralinos using various decay channels are performed using $139\,$fb$^{-1}$ of $pp$ collision data at $\sqrt{s}=13\,$TeV with the ATLAS detector at the Large Hadron Collider.
Searches targeting pure-wino chargino pair production, pure-wino chargino-neutralino production, or higgsino production decaying via Standard Model $W$, $Z$, or $h$ bosons are combined to extend the mass reach to the produced SUSY particles by 30--100\,GeV.
The depth of the sensitivity of the original searches is also improved by the combinations, lowering the 95\% CL cross-section upper limits by 15\%--40\%.
}

\AtlasRefCode{SUSY-2020-05}

\AtlasJournal{PRL}

\PreprintIdNumber{CERN-EP-2024-018}
 
\AtlasJournalRef{Phys. Rev. Lett 133 (2024) 031802}
\AtlasDOI{10.1103/PhysRevLett.133.031802}

\hypersetup{pdftitle={ATLAS document},pdfauthor={The ATLAS Collaboration}}
 
\begin{document}
 
\maketitle

Supersymmetry~\cite{Golfand:1971iw,Volkov:1973ix,Wess:1974tw,Wess:1974jb,Ferrara:1974pu,Salam:1974ig,PhysRevD.3.2415} (SUSY) proposes a superpartner for every Standard Model (SM) particle, where the spin differs by one-half.
It remains one of the more popular beyond the SM theories as it can provide solutions for the hierarchy problem, dark matter, and unification of the fundamental forces~\cite{Sakai:1981gr,Dimopoulos:1981yj,Ibanez:1981yh,Dimopoulos:1981zb}.
Naturalness arguments motivate some SUSY particles to be within reach of the LHC, namely the fermionic superpartners of the gauge and Higgs fields: the charginos $\chinogen$ and neutralinos $\ninogen$~\cite{Barbieri:1987fn,deCarlos:1993yy}.
The lightest neutralino $\ninoone$ (or the gravitino $\gravino$ in general gauge mediated (GGM) SUSY~\cite{Cheung:2007es,Meade:2008wd,Matchev_2000}) is stable in the $R$-Parity~\cite{Farrar:1978xj} conserving scenarios considered here and is an excellent dark matter candidate~\cite{Goldberg:1983nd,Ellis:1983ew}.
In these scenarios, charginos and neutralinos are produced in pairs at the LHC and decay into the $\ninoone$ or $\gravino$ via SM bosons (where the SM boson decays follow SM branching fractions), assuming other SUSY particles are too heavy to play a role.
With the limits on strongly produced SUSY particle masses exceeding $\sim$2$\,\TeV$~\cite{SUSY-2018-22}, electroweakly produced SUSY particles may dominate LHC SUSY production.
Small production cross-sections and decay modes with similar experimental signatures to SM processes make these some of the more challenging searches at the LHC.
 
\begin{figure}[h!]
\centering
\subfloat[][]{\includegraphics[width=0.23\textwidth]{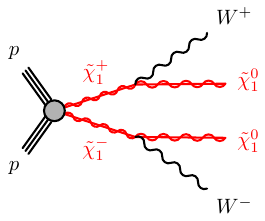}\label{fig:DiagramWW}}      \hspace{0.2cm}
\subfloat[][]{\includegraphics[width=0.23\textwidth]{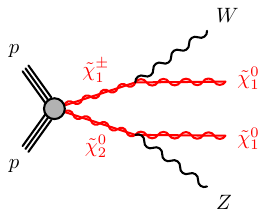}\label{fig:DiagramWZ}}      \hspace{0.2cm}
\subfloat[][]{\includegraphics[width=0.23\textwidth]{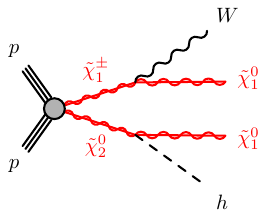}\label{fig:DiagramWh}}      \hspace{0.2cm}
\subfloat[][]{\includegraphics[width=0.23\textwidth]{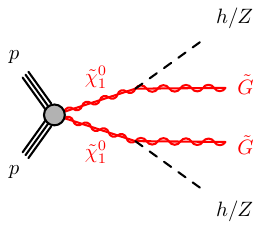}\label{fig:DiagramGGM}}
\caption{Diagrams of the processes in the simplified SUSY models considered in this letter:
(a) wino chargino-pair production decaying via $W$ bosons,
(b) wino chargino--neutralino production decaying via $W$ and $Z$ bosons,
(c) wino chargino--neutralino production decaying via $W$ and $h$ bosons,
and (d) higgsino GGM scenarios. In (d) the $\ninoone$ may be produced via $\chinoonep\chinoonem$, $\chinoonepm\ninoggm$, or $\ninoone\ninotwo$ production.
The grey blob represents all possible intermediate states. For these simplified models, all other SUSY particles are assumed to be heavy and decoupled.
\label{fig:Diagrams}}
\end{figure}

The investigation of electroweakly produced SUSY particles by the ATLAS Collaboration~\cite{PERF-2007-01,ATLAS-TDR-19,PIX-2018-001,ATL-SOFT-PUB-2021-001} comprises searches with multiple final states targeting different production and intermediate decay modes.
These searches are harmonized to allow for the statistical combination of the results, increasing the sensitivity to SUSY by broadening the mass reach and improving the cross-section reach. 
Combining results can be particularly powerful when the searches have different, but complementary, sensitivity to the same SUSY models.
This letter focuses on the pair production of pure-wino or pure-higgsino next-to-lightest SUSY particles (NLSP) decaying into the lightest SUSY particle (LSP) via a SM boson.
The Run 2 electroweak SUSY searches at ATLAS, corresponding to 139\,fb$^{-1}$ of $pp$ LHC collision data at a center-of-mass energy of $\sqrt{s}=13\,\TeV$, are statistically combined for each SUSY scenario shown in Figure~\ref{fig:Diagrams}, as reported in Table~\ref{tab:combis}.
The CMS Collabration have also performed statistical combinations of their electroweak SUSY searches, found in Ref.~\cite{CMS-SUS-21-008}.
 
\begin{table}[ht]
\centering
\caption{The electroweak SUSY production modes considered, along with the multiple decay modes and final states used for the statistical combination.
\label{tab:combis}}
\scriptsize{
\begin{tabular}{ l | c|c|c|c }
\toprule{\bfseries Production mode}                      & Wino                        & Wino                       & Wino                      & Higgsino GGM\\
& $\chinoonep\chinoonem$      &  $\chinoonepm\ninotwo$     &  $\chinoonepm\ninotwo$    & $\chinoonep\chinoonem$, $\chinoonepm\ninoggm$, $\ninoone\ninotwo$ \\
\midrule{\bfseries Decay mode} &                          $\chinoonepm \ra W^\pm\ninoone$ & $\chinoonepm \ra W^\pm\ninoone$& $\chinoonepm \ra W^\pm\ninoone$& $\ninoone \ra  Z/h \gravino$\\
&                            & $\ninotwo \ra Z\ninoone$   & $\ninotwo \ra h\ninoone$   & \\
\midrule{\bfseries Searches} & & & & \\
\emph{All Hadronic}~\cite{SUSY-2018-41}   & \checkmark                 & \checkmark                 & \checkmark                 & \checkmark \\
\emph{1L}~\cite{SUSY-2023-01}             & \checkmark                 & \checkmark                 &                            &        \\
\emph{1Lbb}~\cite{SUSY-2019-08}           &                            &                            & \checkmark                 &         \\
\emph{2L Compressed}~\cite{SUSY-2018-16}  &                            & \checkmark                 &                            &        \\
\emph{2L0J $\Delta m > m(W)$}~\cite{SUSY-2018-32}     & \checkmark                 &                            &                            &        \\
\emph{2L0J $\Delta m \sim m(W)$}~\cite{SUSY-2019-02}     & \checkmark                 &                            &                            &        \\
\emph{2L2J}~\cite{SUSY-2018-05}           &                            & \checkmark                 &                            & \checkmark \\
\emph{2$\tau$}~\cite{SUSY-2023-03}           &                            &                            & \checkmark                 &        \\
\emph{3L}~\cite{SUSY-2019-09}             &                            & \checkmark                 & \checkmark                 &        \\
\emph{SS/3L}~\cite{SUSY-2019-22}          &                            & \checkmark                 & \checkmark                 &        \\
\emph{4L}~\cite{SUSY-2018-02}             &                            &                            &                            & \checkmark \\
\emph{Multi-b}~\cite{SUSY-2020-16}        &                            &                            &                            & \checkmark \\
\bottomrule
\end{tabular}}
\end{table}
 
To obtain the best sensitivity to a new physics signal through a statistical combination of the individual results, the searches used should be statistically independent and not overlap in their event selection for signal regions (SR) or control regions (CR).
Overlap is avoided for the most part by requiring exclusive lepton multiplicity in any search selection, so that 0$\ell$, 1$\ell$, 2$\ell$, 3$\ell$, and 4$\ell$ searches (where $\ell=e, \mu$) are statistically independent.
To achieve this, the searches adopted common loose selection criteria\footnote{Electrons must satisfy $\pt>4.5\,\GeV$, $|\eta|<2.47$, $|z_0\sin\theta|<0.5\,\mathrm{mm}$, and ``LooseAndBLayerLLH'' requirements~\cite{EGAM-2018-01}. Muons must satisfy $\pt>3\,\GeV$, $|\eta|<2.7$, $|z_0\sin\theta|<0.5\,\mathrm{mm}$, and ``Medium'' identification requirements~\cite{MUON-2018-03}.} at the very start of each analysis, allowing the free use of any further criteria without overlapping with other lepton multiplicities.
The \emph{All Hadronic}, \emph{Multi-b}, and \emph{1L} searches found the veto of loose and low-$\pt$ leptons detrimental to signal acceptance. To avoid this, a less stringent veto was adopted,\footnote{Events selected by 0$\ell$ and 1$\ell$ searches must have fewer than three leptons passing the common loose selection, and fewer than two satisfying $\pt > 8 \,\GeV$.} designed to reject events selected by 2$\ell$ or $3\ell$ searches.
The \emph{2L Compressed} search used an even looser muon definition, however, the search selection is unique enough to result in orthogonality to the others used in a combination.
The harmonization procedure was adopted early in the ATLAS Run 2 search programme and proved to be a keystone of this final combination effort.
 
The statistical independence of the searches is verified by inspecting the events selected by SRs and CRs in the data and in high statistics simulation of SUSY signals.
Significant overlaps are observed between those with equal lepton multiplicity selections, e.g. the \emph{All Hadronic} and \emph{Multi-b} searches, and statistical combinations are not performed for those with $>10\%$ overlap.
In these cases, the search with the best expected sensitivity is used and each instance is discussed for the SUSY models in the following.
Otherwise, all SRs used in the combination have zero overlap with other SRs and CRs, while a few CRs have a small $\sim$1\%--2\% overlap with one another.

Limits are set in SUSY simplified models~\cite{Alwall:2008ve,Alwall:2008ag,Alves:2011wf} using a combined profile likelihood fit to the observed yields, the estimate of SM background yields, and the expected SUSY yields in the CRs and SRs.
Systematic uncertainties are included as Gaussian-distributed nuisance parameters in the likelihood fit and can be correlated between CRs and SRs with common nuisance parameters.
The fit parameters are determined by maximizing the product of the Poisson probability functions and the constraints for the nuisance parameters.
The compatibility of a signal scenario with the data observation is assessed by accounting for the SUSY signal in all CRs and SRs scaled by a floating signal normalization factor.
A signal scenario is excluded if the upper limit at 95\% confidence level (CL) of the signal normalization factor obtained in the fit is smaller than that predicted by the cross-section of the scenario~\cite{Read:2002hq}.
Signal cross-sections are calculated to next-to-leading order in the strong coupling constant, adding the resummation of soft gluon emission at next-to-leading-logarithmic accuracy (NLO+NLL)~\cite{Beenakker:1999xh,Debove:2010kf,Fuks:2012qx,Fuks:2013vua,Fiaschi:2018hgm}. The nominal cross-section and the uncertainty are taken from an envelope of cross-section predictions using different parton distribution function sets and factorization and renormalization scales, as described in Ref.~\cite{Borschensky:2014cia}.
 
The statistical combination for each signal scenario is performed with the \textsc{pyhf} package~\cite{pyhf_joss}, using inputs produced by the original search (typically using \textsc{HistFitter}~\cite{Baak:2014wma}), or via the RECAST implementation of the search~\cite{Cranmer_2011}.
The inputs contain information about the yields and uncertainties in the SM background and signal in each CR and SR, as well as the observed data yields.
Systematic uncertainties can be set as correlated between searches, where appropriate, by modifying the inputs to share nuisance parameters in the likelihood fit.
Theory systematic uncertainties in the SM backgrounds and signal are treated as uncorrelated between searches since each search targets a different final state and parameter space.
Experimental systematic uncertainties might be correlated if compatible uncertainty schemes are used by each search to be combined.
However, this is not always possible because the searches to be combined span significant updates in particle reconstruction and identification methods, and the related calibrations, preventing the correlation of multiple sources between searches.
Additionally, incompatible choices for jet systematic schemes were used in individual searches, preventing the correlation of jet energy scale and resolution uncertainties.
Correlating only the allowed sources of experimental systematic uncertainties between searches is found to have a negligible impact on the results.
In this letter, statistical combinations are performed with theory and experimental uncertainties uncorrelated between searches.

\begin{figure}[h!]
\centering
\subfloat[][]{\includegraphics[width=0.45\textwidth]{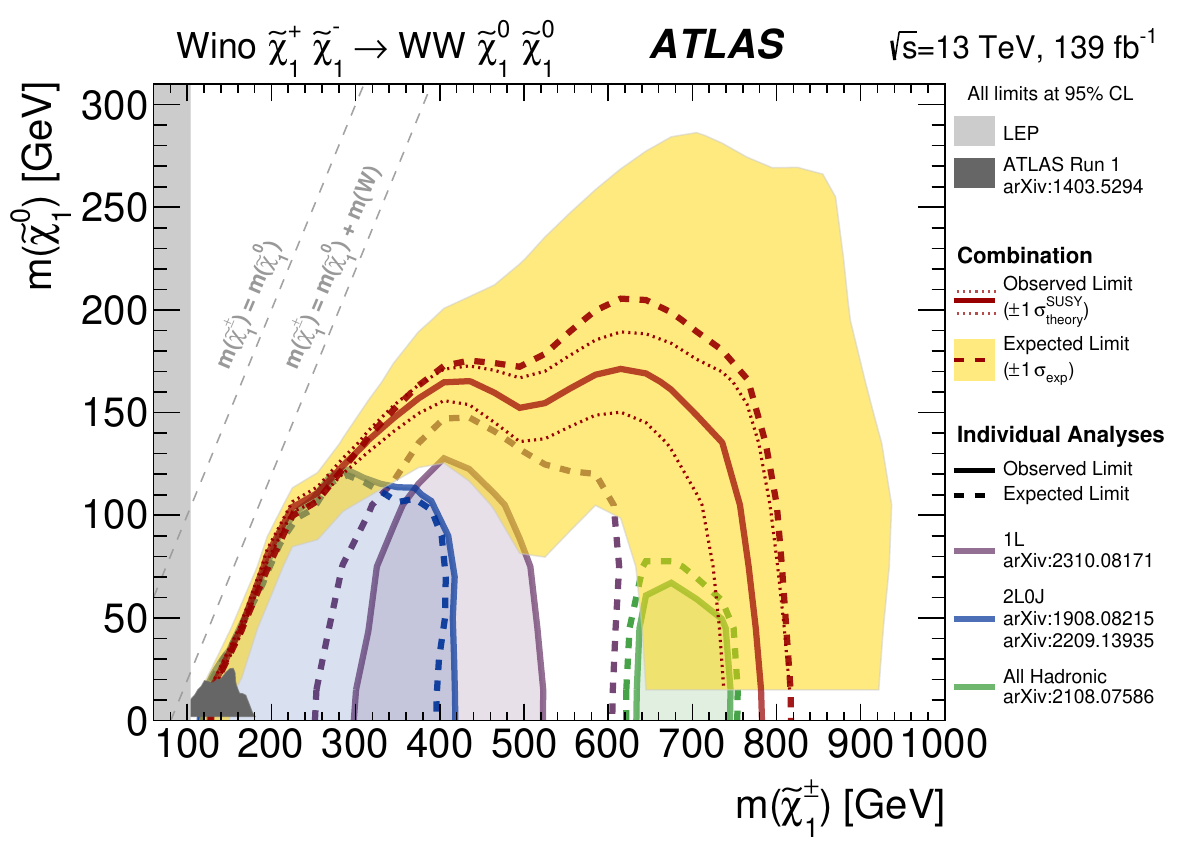}\label{fig:LimitPlotsWW}}
\subfloat[][]{\includegraphics[width=0.45\textwidth]{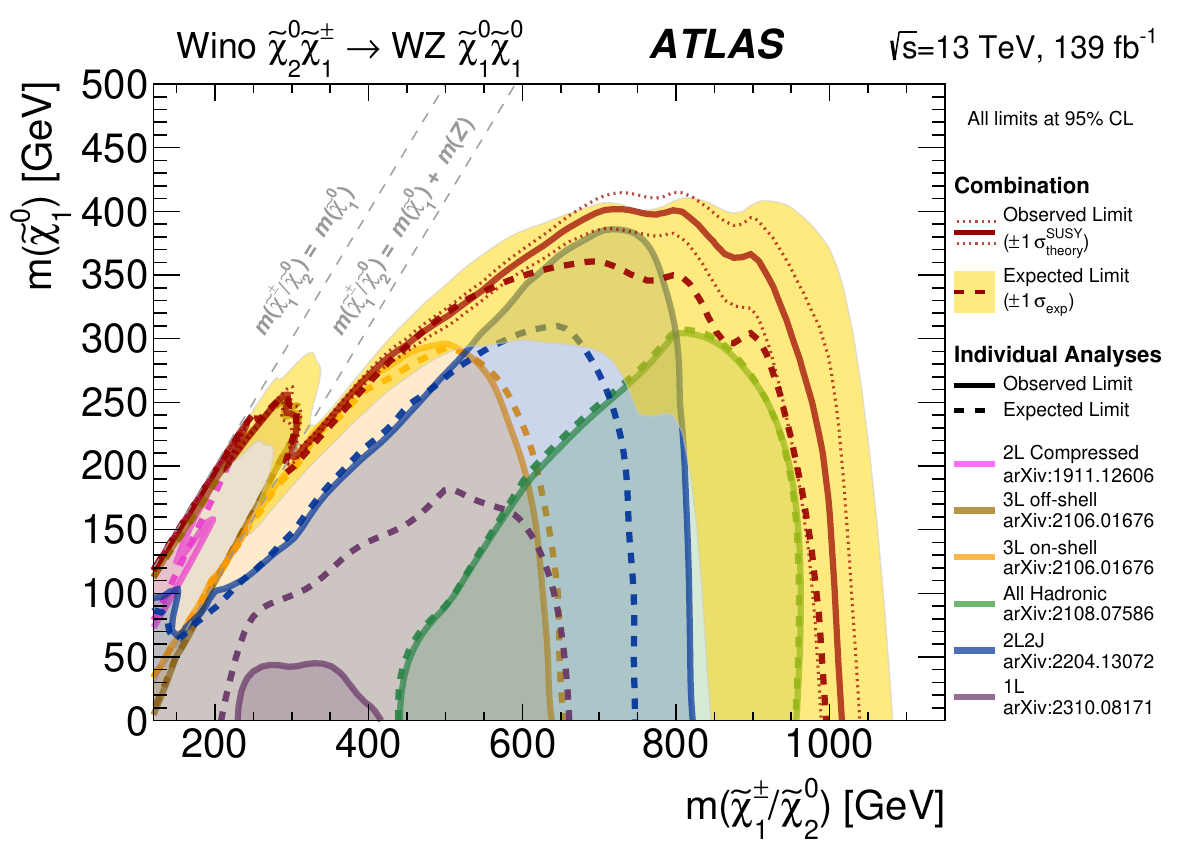}\label{fig:LimitPlotsWZ}}\\
\subfloat[][]{\includegraphics[width=0.45\textwidth]{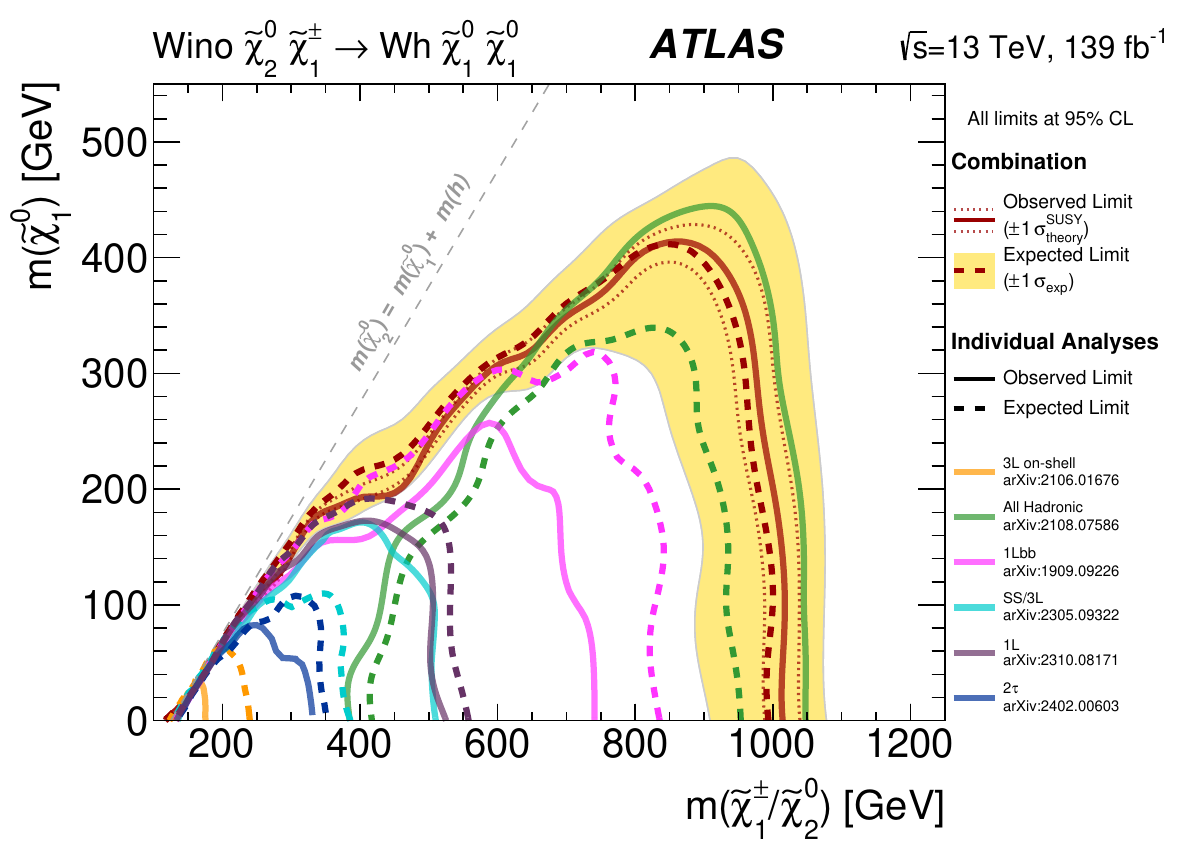}\label{fig:LimitPlotsWh}}
\subfloat[][]{\includegraphics[width=0.45\textwidth]{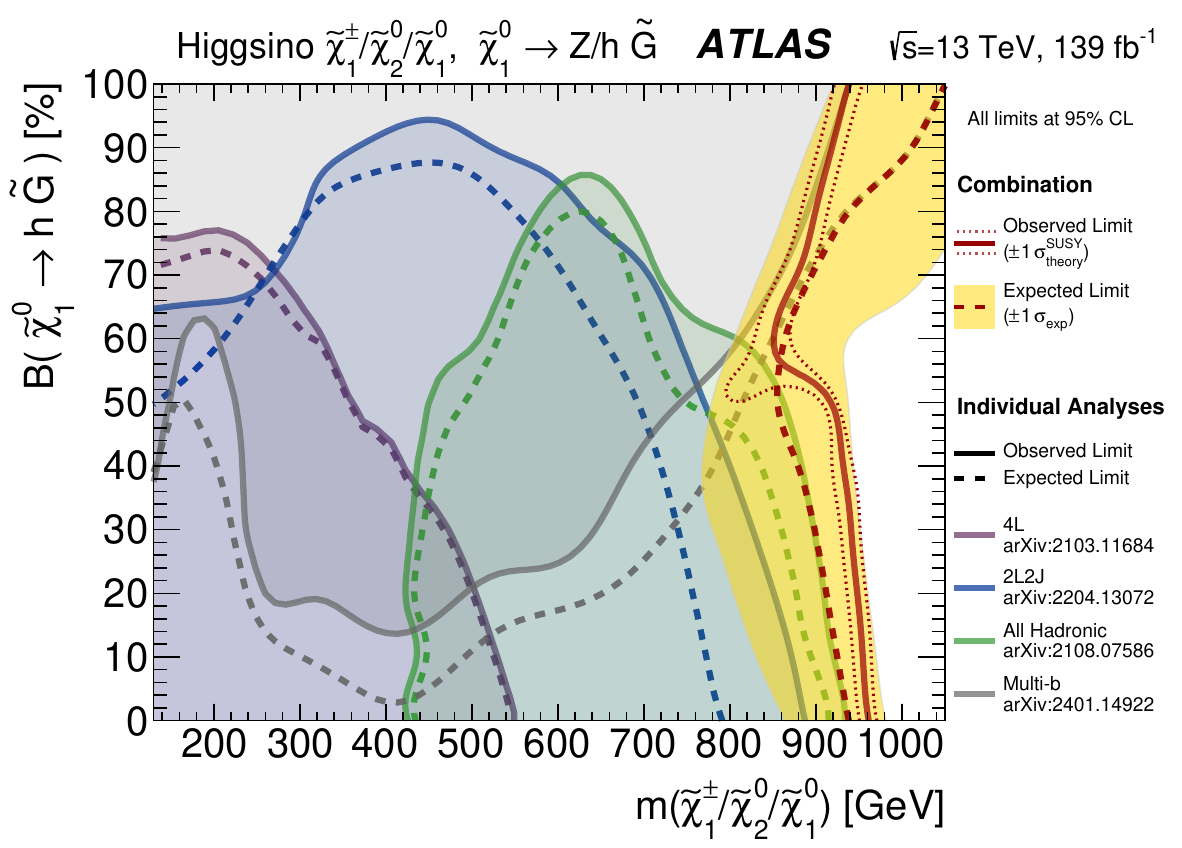}\label{fig:LimitPlotsGGM}}
\caption{The expected (dashed) and observed (solid) 95\% CL exclusion limits on
(a) chargino-pair production decaying via $W$ bosons,
(b) chargino--neutralino production decaying via $W$ and $Z$ bosons,
(c) chargino--neutralino production decaying via $W$ and $h$ bosons,
(d) higgsino GGM scenarios.
The limits are set using a statistical combination of searches targeting each SUSY scenario.
Limits obtained by individual searches are overlaid.
\label{fig:LimitPlots}}
\end{figure}

A simplified model of pure-wino chargino-pair production decaying into $W$ bosons and the LSP 100\% of the time ($\chinoonep\chinoonem$, $\chinoonepm\rightarrow W^{\pm} \ninoone$, as shown in Figure~\ref{fig:DiagramWW}) can produce final states of $\ell\nu\ell\nu\ninoone\ninoone$, $\ell\nu qq\ninoone\ninoone$, or $qqqq\ninoone\ninoone$.
The fully leptonic final state was targeted in two searches: \emph{2L0J $\Delta m > m(W)$} for moderate NLSP-LSP mass splittings and \emph{2L0J $\Delta m \sim m(W)$} for smaller mass splittings. The two \emph{2L0J} searches overlap in their selection, so the search with the lowest expected CL value is used in the statistical combination for each signal scenario.
The semileptonic and fully hadronic final states were targeted by the \emph{1L} and \emph{All Hadronic} searches, respectively, both of which are statistically independent of one another and the \emph{2L0J} searches.
The original exclusion contours in the $m(\chinoonepm)$-$m(\ninoone)$ parameter space are shown in Figure~\ref{fig:LimitPlotsWW}, along with that obtained by the statistical combination of the searches.
The combination of the search results closes the gaps left by the individual searches, and increases the sensitivity to high $\ninoone$ masses, where $\ninoone$ masses are excluded up to $150\,\GeV$ for a $\chinoonepm$ mass of 400--700\,\GeV.
The combination is used to calculate the upper limit on the cross-section for these $\chinoonep\chinoonem$ simplified models, where the limits are improved by 20\%--30\% for $\chinoonepm$ masses of 400--800\,\GeV, compared to the individual searches.
Improvements in the upper limit on the cross-section are particularly important for non-simplified SUSY models where the production cross-section and decay branching fractions may be lower than those in simplified models.\footnote{Non-simplified SUSY models typically describe mixed wino/higgsino/bino charginos and neutralinos.}
 
A second simplified model is considered consisting of pure-wino, mass-degenerate chargino--neutralino pair production decaying into $W$ or $Z$ bosons and the LSP 100\% of the time  ($\chinoonepm\ninotwo$, $\chinoonepm\rightarrow W^{\pm} \ninoone$, $\ninotwo\rightarrow Z \ninoone$, as shown in Figure~\ref{fig:DiagramWZ}).
Searches targeting the fully hadronic, semileptonic, and fully leptonic decays of the SM bosons are considered for a statistical combination, as listed in Table~\ref{tab:combis}, where all searches are statistically independent and can be combined.
The original exclusion contours in the $m(\chinoonepm/\ninotwo)$-$m(\ninoone)$ parameter space are shown in Figure~\ref{fig:LimitPlotsWZ}, along with that obtained by the statistical combination of the searches.
The combination has little impact for small NLSP-LSP mass splittings, where the \emph{2L Compressed} search is uniquely sensitive.
However, at larger mass splittings, multiple searches have common sensitivity and the combination is more effective.
The exclusion contour is extended for high $m(\chinoonepm/\ninotwo)$ by around $50\,\GeV$, while the reach to $m(\ninoone)$ masses is extended by 40--100\,\GeV at $\chinoonepm/\ninotwo$ masses around $550\,\GeV$ and $800\,\GeV$.
The upper limit on the cross-section for these simplified models is improved by 20\%--40\% for $\chinoonepm/\ninotwo$ masses of 600--1000\,\GeV\ relative to respect to the individual searches alone.
 
A third simplified model is considered of pure-wino, mass-degenerate chargino--neutralino pair production decaying into $W$ or Higgs bosons $h$ and the LSP 100\% of the time  ($\chinoonepm\ninotwo$, $\chinoonepm\rightarrow W^{\pm} \ninoone$, $\ninotwo\rightarrow h \ninoone$, as shown in Figure~\ref{fig:DiagramWh}).
The \emph{All Hadronic} and \emph{1Lbb} searches target the $h\rightarrow bb$ decay and dominate the sensitivity to these models, while $h$ decays resulting in leptons are targeted using the \emph{SS/3L}, \emph{3L}, and \emph{2$\tau$} searches and are sensitive to low mass NLSP production.
The \emph{SS/3L} and \emph{3L} searches overlap in their selection, so the search with the lowest expected CL is considered for statistical combination with the other searches for each signal scenario.
The original exclusion contours in the $m(\chinoonepm/\ninotwo)$-$m(\ninoone)$ parameter space are shown in Figure~\ref{fig:LimitPlotsWh}, along with that obtained by the statistical combination of the searches.
The combination smooths out the effects of the small observed deficit seen in the \emph{All Hadronic} search and a small observed excess in the \emph{1Lbb} search, with a stronger expected limit for the combination, but a weaker observed limit than the \emph{All Hadronic} search.
The exclusion contour is extended up to $30\,\GeV$ in $\ninoone$ masses for $\chinoonepm/\ninotwo$ masses of 300--600\,\GeV.
The combination improves the upper limit on the cross-section for these simplified models by 20\%--30\% for $\chinoonepm/\ninotwo$ masses below $600\,\GeV$ compared to the individual searches alone.
 
A fourth simplified model of pure-Higgsino production is considered ($\chinoonep\chinoonem$ / $\chinoonepm\ninoone$ / $\chinoonepm\ninotwo$ / $\ninoone\ninotwo$), the higgsino GGM scenarios, as shown in Figure~\ref{fig:DiagramGGM}. The $\ninotwo$ and $\chinoonepm$ masses are set $1\,\GeV $ above the $\ninoone$ mass to ensure prompt decays.
The $\ninotwo$ and $\chinoonepm$ decay into $\ninoone$ via off-shell $W$ or $Z$ bosons, which in turn decay into unimportant, low momentum ($<1\,\GeV$) final states.
The $\ninoone$ decays into an LSP $\gravino$, either with a $Z$ boson or a $h$ boson.
The higgsino GGM scenarios are parameterized by the mass of the higgsinos and the branching fraction of the $\ninoone$ decay.
These signal scenarios are targeted by the \emph{4L}, \emph{2L2J}, and \emph{All Hadronic} searches selecting leptonic or hadronic decays of the $Z$ boson, and by the \emph{Multi-b} search selecting $h\rightarrow bb$ decays.
The \emph{All Hadronic} and \emph{Multi-b} searches overlap in their selection, so the search with the lowest expected CL is used in the statistical combination.
The original exclusion contours in the $m(\chinoonepm/\ninotwo/\ninoone)$-$\mathcal{B}(\ninoone\rightarrow h \gravino)$ parameter space are shown in Figure~\ref{fig:LimitPlotsGGM}, along with that obtained by the statistical combination of the searches.
Full coverage of the $\ninoone$ branching ratio possibilities is obtained by the individual searches and the combination extends the exclusion by around $60\,\GeV$ for high mass higgsino production.
The upper limit on the cross-section for these simplified models is improved by 15\%--40\% for $\mathcal{B}(\ninoone\rightarrow h \gravino)<80\%$ compared to the individual searches alone.

Statistical combinations of the Run 2 ATLAS electroweak SUSY searches targeting chargino/neutralino production are performed.
Four simplified SUSY models are studied: pure-wino $\chinoonep\chinoonem$ production decaying via $W$ bosons, pure-wino $\chinoonepm\ninotwo$ production decaying via $W$ and $Z$ bosons, pure-wino $\chinoonepm\ninotwo$ production decaying via $W$ and $h$ bosons, and higgsino GGM scenarios.
The combinations extend the sensitivity to SUSY production up to $100\,\GeV$ in NLSP or LSP masses, and the sensitivity to SUSY production cross-sections is increased by up to 40\%.

\FloatBarrier
\section*{Acknowledgements}

% The next lines are included from the .//acknowledgements/Acknowledgements.tex input file

We thank CERN for the very successful operation of the LHC and its injectors, as well as the support staff at
CERN and at our institutions worldwide without whom ATLAS could not be operated efficiently.
 
The crucial computing support from all WLCG partners is acknowledged gratefully, in particular from CERN, the ATLAS Tier-1 facilities at TRIUMF/SFU (Canada), NDGF (Denmark, Norway, Sweden), CC-IN2P3 (France), KIT/GridKA (Germany), INFN-CNAF (Italy), NL-T1 (Netherlands), PIC (Spain), RAL (UK) and BNL (USA), the Tier-2 facilities worldwide and large non-WLCG resource providers. Major contributors of computing resources are listed in Ref.~\cite{ATL-SOFT-PUB-2023-001}.
 
We gratefully acknowledge the support of ANPCyT, Argentina; YerPhI, Armenia; ARC, Australia; BMWFW and FWF, Austria; ANAS, Azerbaijan; CNPq and FAPESP, Brazil; NSERC, NRC and CFI, Canada; CERN; ANID, Chile; CAS, MOST and NSFC, China; Minciencias, Colombia; MEYS CR, Czech Republic; DNRF and DNSRC, Denmark; IN2P3-CNRS and CEA-DRF/IRFU, France; SRNSFG, Georgia; BMBF, HGF and MPG, Germany; GSRI, Greece; RGC and Hong Kong SAR, China; ISF and Benoziyo Center, Israel; INFN, Italy; MEXT and JSPS, Japan; CNRST, Morocco; NWO, Netherlands; RCN, Norway; FCT, Portugal; MNE/IFA, Romania; MESTD, Serbia; MSSR, Slovakia; ARIS and MVZI, Slovenia; DSI/NRF, South Africa; SRC and Wallenberg Foundation, Sweden; SERI, SNSF and Cantons of Bern and Geneva, Switzerland; NSTC, Taipei; TENMAK, T\"urkiye; STFC/UKRI, United Kingdom; DOE and NSF, United States of America.
 
Individual groups and members have received support from BCKDF, CANARIE, CRC and DRAC, Canada; CERN-CZ, PRIMUS 21/SCI/017 and UNCE SCI/013, Czech Republic; COST, ERC, ERDF, Horizon 2020, ICSC-NextGenerationEU and Marie Sk{\l}odowska-Curie Actions, European Union; Investissements d'Avenir Labex, Investissements d'Avenir Idex and ANR, France; DFG and AvH Foundation, Germany; Herakleitos, Thales and Aristeia programmes co-financed by EU-ESF and the Greek NSRF, Greece; BSF-NSF and MINERVA, Israel; Norwegian Financial Mechanism 2014-2021, Norway; NCN and NAWA, Poland; La Caixa Banking Foundation, CERCA Programme Generalitat de Catalunya and PROMETEO and GenT Programmes Generalitat Valenciana, Spain; G\"{o}ran Gustafssons Stiftelse, Sweden; The Royal Society and Leverhulme Trust, United Kingdom.
 
In addition, individual members wish to acknowledge support from Chile: Agencia Nacional de Investigaci\'on y Desarrollo (FONDECYT 1190886, FONDECYT 1210400, FONDECYT 1230812, FONDECYT 1230987); China: National Natural Science Foundation of China (NSFC - 12175119, NSFC 12275265, NSFC-12075060); Czech Republic: PRIMUS Research Programme (PRIMUS/21/SCI/017); EU: H2020 European Research Council (ERC - 101002463); European Union: European Research Council (ERC - 948254), Horizon 2020 Framework Programme (MUCCA - CHIST-ERA-19-XAI-00), European Union, Future Artificial Intelligence Research (FAIR-NextGenerationEU PE00000013), Italian Center for High Performance Computing, Big Data and Quantum Computing (ICSC, NextGenerationEU), Marie Sklodowska-Curie Actions (EU H2020 MSC IF GRANT NO 101033496); France: Agence Nationale de la Recherche (ANR-20-CE31-0013, ANR-21-CE31-0013, ANR-21-CE31-0022, ANR-22-EDIR-0002), Investissements d'Avenir Idex (ANR-11-LABX-0012), Investissements d'Avenir Labex (ANR-11-LABX-0012); Germany: Baden-Württemberg Stiftung (BW Stiftung-Postdoc Eliteprogramme), Deutsche Forschungsgemeinschaft (DFG - CR 312/5-1); Italy: Istituto Nazionale di Fisica Nucleare (FELLINI G.A. n. 754496, ICSC, NextGenerationEU); Japan: Japan Society for the Promotion of Science (JSPS KAKENHI JP21H05085, JSPS KAKENHI JP22H01227, JSPS KAKENHI JP22H04944); Netherlands: Netherlands Organisation for Scientific Research (NWO Veni 2020 - VI.Veni.202.179); Norway: Research Council of Norway (RCN-314472); Poland: Polish National Agency for Academic Exchange (PPN/PPO/2020/1/00002/U/00001), Polish National Science Centre (NCN 2021/42/E/ST2/00350, NCN OPUS nr 2022/47/B/ST2/03059, NCN UMO-2019/34/E/ST2/00393, UMO-2020/37/B/ST2/01043, UMO-2021/40/C/ST2/00187); Slovenia: Slovenian Research Agency (ARIS grant J1-3010); Spain: BBVA Foundation (LEO22-1-603), Generalitat Valenciana (Artemisa, FEDER, IDIFEDER/2018/048), La Caixa Banking Foundation (LCF/BQ/PI20/11760025), Ministry of Science and Innovation (MCIN \& NextGenEU PCI2022-135018-2, MICIN \& FEDER PID2021-125273NB, RYC2019-028510-I, RYC2020-030254-I, RYC2021-031273-I, RYC2022-038164-I), PROMETEO and GenT Programmes Generalitat Valenciana (CIDEGENT/2019/023, CIDEGENT/2019/027); Sweden: Swedish Research Council (VR 2018-00482, VR 2022-03845, VR 2022-04683, VR grant 2021-03651), Knut and Alice Wallenberg Foundation (KAW 2017.0100, KAW 2018.0157, KAW 2018.0458, KAW 2019.0447); Switzerland: Swiss National Science Foundation (SNSF - PCEFP2\_194658); United Kingdom: Leverhulme Trust (Leverhulme Trust RPG-2020-004); United States of America: U.S. Department of Energy (ECA DE-AC02-76SF00515), Neubauer Family Foundation.

% End of text imported from the .//acknowledgements/Acknowledgements.tex input file

\printbibliography
 
\clearpage
% ATLAS Collaboration author list
% Reference date of SUSY-2020-05 is 2023-07-27
% Author list last updated on date 26-JUL-24
% Data extracted on 26-Jul-2024 for paper reference SUSY-2020-05
% at 4:30pm
 
\begin{flushleft}
\hypersetup{urlcolor=black}
{\Large The ATLAS Collaboration}

\bigskip

\AtlasOrcid[0000-0002-6665-4934]{G.~Aad}$^\textrm{\scriptsize 102}$,
\AtlasOrcid[0000-0002-5888-2734]{B.~Abbott}$^\textrm{\scriptsize 120}$,
\AtlasOrcid[0000-0002-1002-1652]{K.~Abeling}$^\textrm{\scriptsize 55}$,
\AtlasOrcid[0000-0001-5763-2760]{N.J.~Abicht}$^\textrm{\scriptsize 49}$,
\AtlasOrcid[0000-0002-8496-9294]{S.H.~Abidi}$^\textrm{\scriptsize 29}$,
\AtlasOrcid[0000-0002-9987-2292]{A.~Aboulhorma}$^\textrm{\scriptsize 35e}$,
\AtlasOrcid[0000-0001-5329-6640]{H.~Abramowicz}$^\textrm{\scriptsize 151}$,
\AtlasOrcid[0000-0002-1599-2896]{H.~Abreu}$^\textrm{\scriptsize 150}$,
\AtlasOrcid[0000-0003-0403-3697]{Y.~Abulaiti}$^\textrm{\scriptsize 117}$,
\AtlasOrcid[0000-0002-8588-9157]{B.S.~Acharya}$^\textrm{\scriptsize 69a,69b,m}$
% ...
\bigskip
\\

% ...
$^{29}$Physics Department, Brookhaven National Laboratory, Upton NY; United States of America.\\
% ...
$^{35}$$^{(a)}$Facult\'e des Sciences Ain Chock, Universit\'e Hassan II de Casablanca;$^{(b)}$Facult\'{e} des Sciences, Universit\'{e} Ibn-Tofail, K\'{e}nitra;$^{(c)}$Facult\'e des Sciences Semlalia, Universit\'e Cadi Ayyad, LPHEA-Marrakech;$^{(d)}$LPMR, Facult\'e des Sciences, Universit\'e Mohamed Premier, Oujda;$^{(e)}$Facult\'e des sciences, Universit\'e Mohammed V, Rabat;$^{(f)}$Institute of Applied Physics, Mohammed VI Polytechnic University, Ben Guerir; Morocco.\\
% ...
$^{49}$Fakult\"{a}t Physik , Technische Universit{\"a}t Dortmund, Dortmund; Germany.\\
% ...
$^{55}$II. Physikalisches Institut, Georg-August-Universit\"{a}t G\"ottingen, G\"ottingen; Germany.\\
% ...
$^{69}$$^{(a)}$INFN Gruppo Collegato di Udine, Sezione di Trieste, Udine;$^{(b)}$ICTP, Trieste;$^{(c)}$Dipartimento Politecnico di Ingegneria e Architettura, Universit\`a di Udine, Udine; Italy.\\
% ...
$^{102}$CPPM, Aix-Marseille Universit\'e, CNRS/IN2P3, Marseille; France.\\
% ...
$^{117}$Department of Physics, New York University, New York NY; United States of America.\\
% ...
$^{120}$Homer L. Dodge Department of Physics and Astronomy, University of Oklahoma, Norman OK; United States of America.\\
% ...
$^{150}$Department of Physics, Technion, Israel Institute of Technology, Haifa; Israel.\\
$^{151}$Raymond and Beverly Sackler School of Physics and Astronomy, Tel Aviv University, Tel Aviv; Israel.\\
% ...
$^{a}$ Also Affiliated with an institute covered by a cooperation agreement with CERN.\\
$^{b}$ Also at An-Najah National University, Nablus; Palestine.\\
$^{c}$ Also at Borough of Manhattan Community College, City University of New York, New York NY; United States of America.\\
$^{d}$ Also at Center for High Energy Physics, Peking University; China.\\
$^{e}$ Also at Center for Interdisciplinary Research and Innovation (CIRI-AUTH), Thessaloniki; Greece.\\
$^{f}$ Also at Centro Studi e Ricerche Enrico Fermi; Italy.\\
% ...
$^{m}$ Also at Department of Physics, King's College London, London; United Kingdom.\\
% ...

\end{flushleft}

% Created with Glance <Atlas.Glance@cern.ch>

\end{document}